\documentclass[aps,pra,twocolumn,showpacs,preprintnumbers,amsmath,amssymb]{revtex4-1}

\usepackage{graphicx}
\usepackage{amsmath}
\usepackage{amssymb}
\usepackage{bm}
\usepackage{braket}

\begin{document}

\title{Fidelity Dynamics of a Jaynes--Cummings Qubit under Markovian and Non-Markovian Coupling Noise}

\author{Fang Wang}
\affiliation{%
	Department of Physics, Stevens Institute of Technology, Hoboken, New Jersey 07030, USA\\
}%
\author{Da-Wei Luo}
\affiliation{%
	Department of Physics, Stevens Institute of Technology, Hoboken, New Jersey 07030, USA\\
}%
\author{Ting Yu}
\email{tyu1@stevens.edu}
\affiliation{%
	Department of Physics, Stevens Institute of Technology, Hoboken, New Jersey 07030, USA\\
}%

\date{\today}

\begin{abstract}
We analytically investigate the fidelity dynamics of a Jaynes--Cummings qubit coupled to a single-mode cavity in the presence of stochastic fluctuations of the qubit--cavity coupling. Such fluctuations naturally arise in cavity- and circuit-QED platforms due to fabrication imperfections, control-field noise, mechanical vibrations, and environmental disturbances. Since the qubit--cavity coupling governs the rate of coherent energy and information exchange, understanding the effects of its fluctuations is essential for assessing the performance of quantum communication, quantum sensing, and quantum information-processing protocols. We consider both Gaussian white noise and Ornstein--Uhlenbeck noise with finite correlation time as models for the coupling fluctuations. Exploiting the closed dynamics within each excitation manifold, we derive exact analytical expressions for the noise-averaged density matrix and the fidelity relative to the ideal noiseless evolution. We show that the fidelity is determined by a single decoherence factor governed by the integrated noise correlation function, establishing a direct connection between microscopic coupling fluctuations and macroscopic coherence loss. In the white-noise limit, the fidelity decays exponentially, whereas finite noise correlation times produce nonexponential decay and slower coherence loss. We further show that the crossover between Markovian-like and non-Markovian regimes is controlled by the ratio of the noise correlation rate to the coherent coupling strength, and that decoherence increases with photon number because of the enhanced effective qubit--cavity coupling. These results provide a unified analytical framework for quantifying decoherence induced by stochastic coupling fluctuations and for evaluating the robustness of cavity-QED protocols under correlated noise.
\end{abstract}


\maketitle

\section{Introduction}

The Jaynes--Cummings (JC) model is one of the most fundamental models in quantum optics, describing the coherent interaction between a two-level system and a single quantized electromagnetic mode \cite{jaynes1963comparison}. Owing to its exact solvability, the JC model has provided a cornerstone for understanding light--matter interactions, predicting phenomena such as vacuum Rabi oscillations, collapse-and-revival dynamics, entanglement generation, and nonclassical states of light \cite{sanchez1983theory,narozhny1981coherence,rempe1987observation,shore1993jaynes,buzek1995quantum}. Experimental demonstrations in cavity and circuit quantum electrodynamics (QED) have firmly established the JC model as a versatile platform for coherent quantum control and quantum information processing \cite{raimond2001manipulating,mabuchi2002cavity,wallraff2004strong,blais2021circuit,brune1996quantum}. More recently, extensions of the JC framework have continued to play important roles in ultrastrong light--matter coupling, quantum thermodynamics, quantum reservoir computing, and distributed quantum architectures \cite{forn2019ultrastrong,frisk2019ultrastrong,cirac1997quantum,das2025quantum}. Despite these advances, realistic cavity- and circuit-QED systems are inevitably open quantum systems whose dynamics are influenced by environmental fluctuations and dissipation. Open quantum systems theory therefore provides the fundamental framework for describing decoherence and its impact on coherent quantum evolution \cite{breuer2002theory,de2017dynamics}. Since decoherence remains one of the primary obstacles to scalable quantum technologies, understanding and mitigating environmental noise has become a central topic in quantum information science.

While the Markovian approximation provides an accurate description for quantum systems weakly coupled to large memoryless reservoirs, many realistic environments exhibit finite correlation times that invalidate this assumption \cite{rivas2014quantum,breuer2016colloquium,de2017dynamics}.The system evolution depends not only on its instantaneous state but also on its previous history, giving rise to non-Markovian dynamics. The presence of environmental memory can significantly modify dissipative processes, leading to nonexponential relaxation, coherence revival, entanglement recovery, and information backflow from the environment to the system \cite{breuer2009measure}. These features have been observed across cavity-QED systems~\cite{kaer2010non,guimond2016rabi},
superconducting circuits~\cite{malekakhlagh2016non,gulacsi2023signatures,odeh2025non,white2020demonstration},
spin and qubit systems~\cite{strunz2004convolutionless,yu2004non,jing2010non,zhao2011dynamics,jing2013many,shi2013non},
and photonic reservoirs~\cite{guimond2016rabi},
demonstrating that non-Markovian effects are ubiquitous rather than exceptional in realistic quantum devices.

Considerable effort has been devoted to non-Markovian open quantum systems. Among the most successful approaches are non-Markovian quantum state diffusion (QSD) and stochastic Schrodinger equation methods, which provide powerful tools for treating colored-noise environments with finite memory \cite{diosi1997non,strunz1999open,yu1999non,yu2004finite}. These methods have enabled accurate descriptions of open quantum dynamics beyond the Born--Markov approximation, revealing that environmental memory can slow decoherence, preserve entanglement, induce coherence revivals, and enhance quantum control under suitable conditions.\cite{yu1999non,yu2004non,chen2014exact}. 
Environmental memory can improve quantum control strategies, enhance parameter estimation and quantum metrology, and assist in protecting fragile quantum coherence and entanglement against environmental degradation \cite{vasile2011continuous,bylicka2014non,li2018concepts}. Understanding the crossover between Markovian and non-Markovian regimes is therefore of both fundamental interest and practical importance for the design of robust quantum information processing devices. Among the various models of non-Markovian environments, colored Gaussian noise with finite correlation time provides a particularly convenient description because it captures essential memory effects while remaining analytically tractable.


To characterize temporally correlated fluctuations, the Ornstein--Uhlenbeck (OU) process provides one of the most widely used colored-noise models because it incorporates a finite correlation time while remaining analytically tractable \cite{gardiner2004quantum}. The OU process naturally interpolates between strongly correlated and white-noise limits and has been extensively employed in the theory of non-Markovian quantum dynamics and stochastic Schrodinger equations \cite{diosi1997non,strunz1999open, ma2014crossover}. Since the coupling strength directly determines coherent excitation exchange between the qubit and cavity, stochastic fluctuations can significantly modify Rabi oscillations and coherent state evolution, particularly in superconducting circuit-QED platforms operating in the strong-coupling regime \cite{koch2007charge,krantz2019quantum, de2013magnetization}. More recently, colored-noise descriptions have been used to investigate fidelity degradation and decoherence in noisy qubit systems, highlighting the important role played by finite environmental memory in preserving quantum coherence \cite{de2024qubit, yu2004non, chen2014exact}. These observations motivate a systematic analysis of stochastic coupling fluctuations within the JC framework.

The performance of quantum devices is commonly evaluated using quantum fidelity, which quantifies the similarity between an actual quantum state or process and its ideal counterpart \cite{nielsen2010quantum,nielsen2002simple,gilchrist2005distance}. Fidelity has become one of the standard benchmarks for quantum gates, state-transfer protocols, and quantum sensing because it characterizes the accumulation of noise and operational errors\cite{liang2019quantum, gu2010fidelity, pedersen2007fidelity}. In cavity- and circuit-QED systems, maintaining high fidelity is essential for reliable quantum-state preparation and manipulation \cite{degen2017quantum}. In our work, since the JC dynamics depend explicitly on the excitation number, the influence of stochastic coupling fluctuations is expected to become increasingly pronounced at higher photon numbers. Understanding this photon-number dependence is therefore important for assessing the robustness of quantum protocols beyond the single-excitation regime.

In this work, we analytically investigate the fidelity dynamics of a single qubit coupled to a single-mode cavity field through a stochastic coupling strength
\begin{equation}
g(t)=g_0+\eta(t),
\end{equation}
where $\eta(t)$ is modeled as either Gaussian white noise or Ornstein--Uhlenbeck noise. By exploiting the closed two-dimensional dynamics within each JC excitation manifold, we derive exact expressions for the noise-averaged density matrix and the fidelity relative to the ideal noiseless evolution. We show that the fidelity is governed by a single decoherence factor determined by the integrated noise correlation function. This compact result enables a direct comparison between Markovian and non-Markovian noise regimes and reveals how finite correlation times can substantially suppress decoherence compared with white-noise fluctuations. We further demonstrate that the crossover between these regimes is controlled by the ratio of the noise correlation rate to the coherent coupling strength and that decoherence is enhanced with increasing photon number through the characteristic $(n+1)$ scaling of the JC interaction.

\section{Model and Hamiltonian}

\begin{figure}[htbp]
    \centering
    \includegraphics[width=0.56\textwidth]{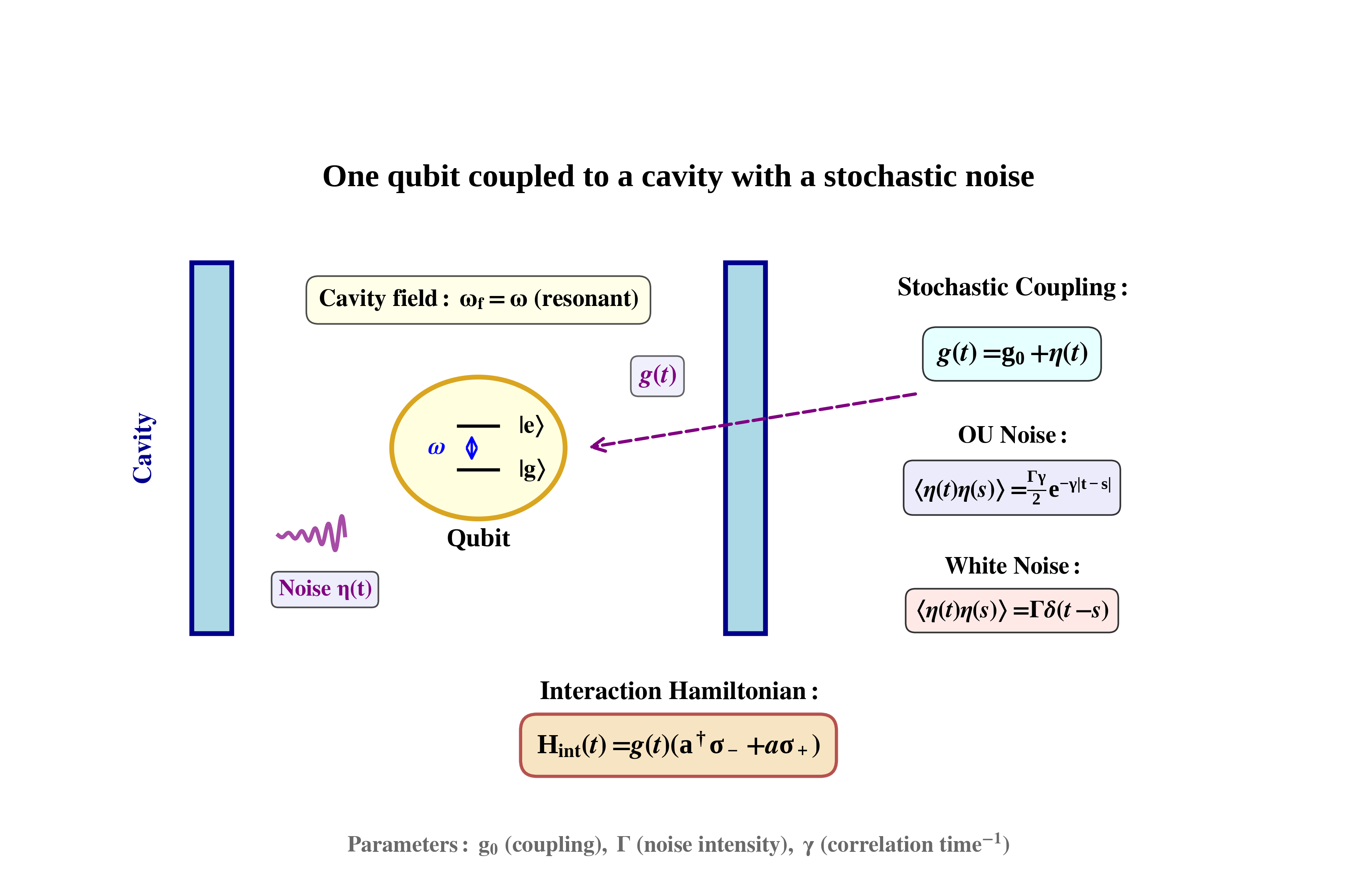}
    \caption{Schematic diagram of a two-level qubit coupled to a single-mode cavity field with stochastic coupling fluctuations. The coupling strength $g(t) = g_0 + \eta(t)$ includes a constant term $g_0$ and a stochastic noise term $\eta(t)$, which can be modeled as either Ornstein-Uhlenbeck (OU) noise or white noise. The system is studied under the resonant condition $\omega_f = \omega$, where the qubit transition frequency matches the cavity mode frequency.}
    \label{fig:cavity_qed_model}
\end{figure}

We consider a two-level system (qubit) interacting with a single-mode cavity field as shown in Fig.~\ref{fig:cavity_qed_model}. The total Hamiltonian in the Schr\"odinger picture is
\begin{equation}
H_{\mathrm{tot}}(t) = \frac{\hbar\omega}{2} \sigma_{z} + \hbar\omega_{f} a^{\dagger} a + \hbar g(t) \left( a^{\dagger} \sigma_{-} + a \sigma_{+} \right),
\label{eq:total_hamiltonian}
\end{equation}
where $\omega$ is the qubit transition frequency, $\sigma_z$ is the Pauli-$z$ operator, and $\sigma_{\pm}$ are the raising and lowering operators. The operators $a^\dagger$ and $a$ are the bosonic creation and annihilation operators of the cavity mode with frequency $\omega_f$. The coupling strength $g(t)$ is a stochastic noise coupling. Throughout this work, we adopt natural units where $\hbar = 1$, and therefore omit $\hbar$ in all subsequent expressions. In these units, the Hamiltonian simplifies to Eq.~\eqref{eq:total_hamiltonian} without the explicit $\hbar$ factors. We focus on the resonant case $\omega_f = \omega$, where maximal energy exchange occurs. Moving to the interaction picture with respect to the free Hamiltonian
\begin{equation}
H_0 = \frac{\omega}{2} \sigma_{z} + \omega a^{\dagger} a,
\end{equation}
and applying the rotating-wave approximation (valid for $g(t) \ll \omega$), the interaction Hamiltonian becomes
\begin{equation}
H_{\mathrm{int}}(t) = g(t) \left( a^{\dagger} \sigma_{-} + a \sigma_{+} \right).
\label{eq:interaction_hamiltonian}
\end{equation}
The coupling strength is modeled as
\begin{equation}
g(t) = g_0 + \eta(t),
\label{eq:coupling}
\end{equation}
where $g_0$ is the constant coupling strength and $\eta(t)$ is a stochastic noise term. We consider two noise models:

\textbf{Ornstein-Uhlenbeck (OU) Noise:} The noise correlation function is
\begin{equation}
\mathcal{M}[\eta(t) \eta(s)] = \frac{\Gamma \gamma}{2} e^{-\gamma |t - s|},
\label{eq:ou_correlation}
\end{equation}
where $\Gamma$ is the noise intensity, $\gamma$ is the inverse correlation time, and $\mathcal{M}[\cdot]$ denotes the ensemble average with $\mathcal{M}[\eta(t)] = 0$.

\textbf{White Noise:} In the limit $\gamma \rightarrow \infty$, the OU noise reduces to white noise:
\begin{equation}
\mathcal{M}[\eta(t) \eta(s)] = \Gamma \delta(t - s).
\label{eq:white_correlation}
\end{equation}

\section{Time Evolution}

The time evolution operator in the interaction picture is given by
\begin{equation}
U(t) = \mathcal{T} \exp\left[ -i \int_0^t H_{\mathrm{int}}(\tau) \, d\tau \right],
\label{eq:evolution_operator}
\end{equation}
where $\mathcal{T}$ is the time-ordering operator. Since $H_{\mathrm{int}}(t)$ and $H_{\mathrm{int}}(s)$ commute for all $t$ and $s$ (as $g(t)$ is a scalar function), we can drop the time-ordering. Define the effective coupling parameter:
\begin{equation}
G(t) = \int_0^t g(\tau) \, d\tau = g_0 t + \int_0^t \eta(\tau) \, d\tau.
\label{eq:effective_coupling}
\end{equation}

Then the evolution operator becomes
\begin{equation}
U(t) = \exp\left[ -i G(t) \left( a^{\dagger} \sigma_{-} + a \sigma_{+} \right) \right].
\label{eq:U_G}
\end{equation}

The JC model interaction couples pairs of states with the same total excitation number. For a system with $n$ photons in the cavity, the interaction couples the manifold $\{\ket{0, n+1}, \ket{1, n}\}$, where the first index denotes the qubit state (0 for ground, 1 for excited) and the second denotes the photon number. We consider an initial state of the form
\begin{equation}
\ket{\psi_0} = \alpha \ket{0, n+1} + \beta \ket{1, n},
\label{eq:initial_state_manifold}
\end{equation}
where $\alpha$ and $\beta$ are complex coefficients with $|\alpha|^2 + |\beta|^2 = 1$. This represents the most general normalized state within this two-dimensional JC manifold. The evolution under $H_{\mathrm{int}}(t)$ remains confined to this subspace.

The interaction Hamiltonian in this subspace can be written in matrix form:
\begin{small}
\begin{equation}
H_{\mathrm{int}}(t) = g(t) \sqrt{n+1} 
\begin{pmatrix}
0 & 1 \\
1 & 0
\end{pmatrix} = g(t) \sqrt{n+1} \, \sigma_x,
\label{eq:H_matrix}
\end{equation}
\end{small}
where $\sigma_x$ is the Pauli-$x$ matrix in the $\{\ket{0, n+1}, \ket{1, n}\}$ basis.

The evolution operator in this subspace is
\begin{equation}
U(t) = 
\begin{pmatrix}
\cos(\Omega) & -i \sin(\Omega) \\
-i \sin(\Omega) & \cos(\Omega)
\end{pmatrix},
\label{eq:U_explicit}
\end{equation}
where $\Omega = G(t)\sqrt{n+1}$.

For the initial state given in Eq.~\eqref{eq:initial_state_manifold}, the evolved state is:
\begin{equation}
\begin{split}
\ket{\psi(t)} = & U(t) \ket{\psi_0}\\
=&\left[\alpha \cos(\Omega) - i \beta \sin(\Omega)\right] \ket{0, n+1} \\
& + \left[\beta \cos(\Omega) - i \alpha \sin(\Omega)\right] \ket{1, n}.
\end{split}
\label{eq:evolved_state}
\end{equation}
Due to the stochastic nature of $g(t)$, we must perform ensemble averaging. The density matrix of the system is:
\begin{equation}
\rho(t) = \mathcal{M}[U(t) \ket{\psi_0}\bra{\psi_0} U^\dagger(t)].
\label{eq:density_matrix}
\end{equation}

Define the decoherence factor:
\begin{equation}
D(t) = e^{-(n+1)\mathcal{M}[\xi^2(t)]/2}.
\label{eq:decoherence_factor}
\end{equation}

For OU noise:
\begin{equation}
D_{\mathrm{OU}}(t) = \exp\left[-\frac{(n+1)\Gamma}{2\gamma}\left(\gamma t - 1 + e^{-\gamma t}\right)\right].
\label{eq:D_OU}
\end{equation}

For white noise:
\begin{equation}
D_{\mathrm{white}}(t) = e^{-(n+1)\Gamma t/2}.
\label{eq:D_white}
\end{equation}

Working in the two-dimensional JC manifold spanned by
$\{|0,n+1\rangle, |1,n\rangle\}$, we write
\begin{equation}
|\psi(t)\rangle =
c_0(t)\,|0,n+1\rangle + c_1(t)\,|1,n\rangle,
\end{equation}
with amplitudes
\begin{align}
c_0(t) &= \alpha \cos\Omega - i \beta \sin\Omega, \\
c_1(t) &= \beta \cos\Omega - i \alpha \sin\Omega,
\end{align}
where $\Omega = G(t)\sqrt{n+1}$ and $G(t)$ is defined in
Eq.~\eqref{eq:effective_coupling}.
The density matrix in this basis is
\begin{small}
\begin{align}
& \rho(t) =
\begin{pmatrix}
\rho_{00}(t) & \rho_{01}(t) \\
\rho_{10}(t) & \rho_{11}(t)
\end{pmatrix}, \\
& \rho_{ij}(t) = \mathcal{M}\!\big[c_i(t)c_j^*(t)\big].
\end{align}
\end{small}

It is convenient to separate the deterministic and stochastic parts of
$\Omega(t)$. Using
\begin{small}
\begin{equation}
\Omega(t) = \sqrt{n+1}\,G(t) = \Omega_0(t) + \sqrt{n+1}\,\xi(t),
\end{equation}
with
\begin{align}
&\Omega_0(t) = g_0 t\sqrt{n+1}, \nonumber \\
&\xi(t) = \int_0^t \eta(\tau)\,d\tau,
\end{align}
\end{small}
and the variance $\mathcal{M}[\xi^2(t)]$ given in
Eqs.~\eqref{eq:xi_variance_OU}--\eqref{eq:xi_variance_white}, the
decoherence factor defined in Eq.~\eqref{eq:decoherence_factor} reads
\begin{equation}
D(t) = \exp\!\left[-\frac{(n+1)}{2}\,\mathcal{M}\big[\xi^2(t)\big]\right].
\end{equation}

Define
\begin{align}
\Delta &= |\alpha|^2 - |\beta|^2, \nonumber \\
R &= \Re(\alpha\beta^*), \nonumber \\
I &= \Im(\alpha\beta^*),
\end{align}
and introduce the shorthand
\begin{small}
\begin{align}
C(t) &= \cos\big(2\Omega_0(t)\big)\,D^4(t), \\
S(t) &= \sin\big(2\Omega_0(t)\big)\,D^4(t).
\end{align}
\end{small}
Assuming the initial qubit state is normalized,
\begin{footnotesize}
\begin{align}
\rho_{00}(t)
&= \frac{1}{2}
  + \frac{\Delta}{2} C(t)
  - I\, S(t), \label{eq:rho00_final}\\[2mm]
\rho_{11}(t)
&= \frac{1}{2}
  - \frac{\Delta}{2} C(t)
  + I\, S(t), \label{eq:rho11_final}\\[2mm]
\rho_{01}(t)
&= R + i I\, C(t)
   + i\frac{\Delta}{2} S(t), \label{eq:rho01_final}\\[2mm]
\rho_{10}(t) &= \rho_{01}^*(t). \label{eq:rho10_final}
\end{align}
\end{footnotesize}

Thus, the noise-averaged density matrix in the
$\{|0,n+1\rangle,|1,n\rangle\}$ manifold takes the compact form
\begin{small}
\begin{equation}
\rho(t) =
\begin{pmatrix}
\frac{1}{2} + \frac{\Delta}{2} C(t) - I S(t)
&
R + i I C(t) + i\frac{\Delta}{2} S(t) \\[1mm]
R - i I C(t) - i\frac{\Delta}{2} S(t)
&
\frac{1}{2} - \frac{\Delta}{2} C(t) + I S(t)
\end{pmatrix}.
\label{eq:rho_matrix_final}
\end{equation}
\end{small}

The detailed derivation can be found in the Appendix~\ref{app:derivation}.

\subsection{Reduced qubit state}

If we are interested only in the qubit, we trace out the cavity mode.
In the present single-manifold description, the reduced qubit density
matrix is diagonal in the energy basis because
$\langle n+1|n\rangle = 0$:
\begin{equation}
\rho_{\mathrm{q}}(t) = \mathrm{Tr}_{\mathrm{cav}}[\rho(t)]
= \rho_{00}(t)\,|0\rangle\langle 0|
 + \rho_{11}(t)\,|1\rangle\langle 1|.
\label{eq:rho_qubit}
\end{equation}
The populations $\rho_{00}(t)$ and $\rho_{11}(t)$ follow the
stochastically damped JC oscillations governed by the
decoherence factors $C(t)$ and $S(t)$, which are fully specified
by $D(t)$ for OU and white noise.

\section{Fidelity Analysis}
The fidelity quantifies how well the actual evolved state matches a target state, providing a measure of the quality of quantum operations in the presence of noise. We analyze the fidelity between the noise-averaged density matrix and the ideal noiseless evolution.

For a mixed state described by density matrix $\rho$ and a pure target state $\ket{\psi_{\mathrm{target}}}$, the fidelity is defined as
\begin{equation}
F = \bra{\psi_{\mathrm{target}}} \rho \ket{\psi_{\mathrm{target}}}.
\label{eq:fidelity_def}
\end{equation}

In our case, we compare the noise-averaged state $\rho(t)$ with the ideal state that would result from deterministic coupling $g_0$ (i.e., in the absence of stochastic fluctuations, $\eta(t) = 0$).

\subsection{Ideal Target State}

For purely deterministic evolution with coupling strength $g_0$, the evolution operator is
\begin{equation}
U_0(t) = \exp\left[-i g_0 t \sqrt{n+1} \, \sigma_x \right],
\end{equation}
and the ideal evolved state is
\begin{equation}
\ket{\psi_{\mathrm{ideal}}(t)} = U_0(t) \ket{\psi_0},
\label{eq:ideal_state}
\end{equation}
where $\ket{\psi_0} = \alpha \ket{0, n+1} + \beta \ket{1, n}$ is the initial state defined in Eq.~\eqref{eq:initial_state_manifold}.

Let $\Omega_0(t) = g_0 t \sqrt{n+1}$. Then the ideal state is
\begin{equation}
\begin{split}
\ket{\psi_{\mathrm{ideal}}(t)} 
&= \left[\alpha \cos(\Omega_0) - i \beta \sin(\Omega_0)\right] \ket{0, n+1} \\
& \quad + \left[\beta \cos(\Omega_0) - i \alpha \sin(\Omega_0)\right] \ket{1, n}.
\end{split}
\label{eq:ideal_evolved}
\end{equation}

\subsection{Fidelity Calculation}

The fidelity between the noise-averaged density matrix $\rho(t)$ and the ideal state is
\begin{equation}
F(t) = \bra{\psi_{\mathrm{ideal}}(t)} \rho(t) \ket{\psi_{\mathrm{ideal}}(t)}.
\label{eq:fidelity}
\end{equation}

In the $\{\ket{0, n+1}, \ket{1, n}\}$ basis, the ideal state has components
\begin{align}
c_0^{\mathrm{ideal}} &= \alpha \cos(\Omega_0) - i \beta \sin(\Omega_0), \\
c_1^{\mathrm{ideal}} &= \beta \cos(\Omega_0) - i \alpha \sin(\Omega_0).
\end{align}

The fidelity can be written as
\begin{equation}
F(t) = \sum_{i,j=0}^{1} (c_i^{\mathrm{ideal}})^* \, \rho_{ij}(t) \, c_j^{\mathrm{ideal}}.
\label{eq:fidelity_expanded}
\end{equation}

Expanding this expression:
\begin{equation}
\begin{split}
F(t) &= |c_0^{\mathrm{ideal}}|^2 \rho_{00}(t) + |c_1^{\mathrm{ideal}}|^2 \rho_{11}(t) \\
&\quad + (c_0^{\mathrm{ideal}})^* c_1^{\mathrm{ideal}} \, \rho_{01}(t)
       + (c_1^{\mathrm{ideal}})^* c_0^{\mathrm{ideal}} \, \rho_{10}(t).
\end{split}
\label{eq:fidelity_terms}
\end{equation}

Using the notation $C(t) = \cos(2\Omega_0)D^4(t)$ and $S(t) = \sin(2\Omega_0)D^4(t)$, we obtain:
\begin{footnotesize}
\begin{equation}
\begin{split}
F(t) &= |c_0^{\mathrm{ideal}}|^2 \left[\frac{1}{2} + \frac{\Delta}{2} C(t) - I S(t)\right] \\
&\quad + |c_1^{\mathrm{ideal}}|^2 \left[\frac{1}{2} - \frac{\Delta}{2} C(t) + I S(t)\right] \\
&\quad + \left[R - i I \cos(2\Omega_0) - i \frac{\Delta}{2} \sin(2\Omega_0)\right]
         \!\left[R + i I C(t) + i\frac{\Delta}{2} S(t)\right] \\
&\quad + \left[R + i I \cos(2\Omega_0) + i \frac{\Delta}{2} \sin(2\Omega_0)\right]
         \!\left[R - i I C(t) - i\frac{\Delta}{2} S(t)\right].
\end{split}
\label{eq:fidelity_full}
\end{equation}
\end{footnotesize}

Using the normalization condition $|c_0^{\mathrm{ideal}}|^2 + |c_1^{\mathrm{ideal}}|^2 = 1$
and noting that $|c_0^{\mathrm{ideal}}|^2 - |c_1^{\mathrm{ideal}}|^2 = \Delta\cos(2\Omega_0) - 2I\sin(2\Omega_0)$,
the cross-terms proportional to $I\Delta$ cancel between the diagonal and off-diagonal contributions. Collecting the remaining terms and using $\cos(2\Omega_0)\,C(t) + \sin(2\Omega_0)\,S(t) = D^4(t)$ together with $\Delta^2/2 + 2I^2 = 1/2 - 2R^2$ (which follows from normalization and $|\alpha\beta^*|^2 = R^2 + I^2$), the fidelity simplifies to:
\begin{equation}
F(t) = \frac{1 + D^4(t)}{2} + 2R^2 \big(1 - D^4(t)\big),
\label{eq:fidelity_compact}
\end{equation}
where $R = \Re(\alpha\beta^*)$.

\subsection{Special Cases}

\subsubsection{Equal Superposition: $|\alpha| = |\beta| = 1/\sqrt{2}$ (real coefficients)}

For an initially balanced superposition with real coefficients $\alpha = \beta = 1/\sqrt{2}$,
we have $R = \Re(\alpha\beta^*) = 1/2$, so $2R^2 = 1/2$ and the fidelity becomes:
\begin{equation}
F(t) = \frac{1 + D^4(t)}{2} + \frac{1 - D^4(t)}{2} = 1.
\label{eq:fidelity_equal}
\end{equation}

The fidelity remains unity for this specific equal superposition with real coefficients. This reflects a symmetry of the model in which noise-induced phase fluctuations do not affect the overlap with the ideal state. More generally, $F = 1$ whenever $|\Re(\alpha\beta^*)| = 1/2$. Note that a superposition with a relative phase, e.g.\ $\alpha = 1/\sqrt{2}$, $\beta = i/\sqrt{2}$, has $R = 0$ and thus $F(t) = [1+D^4(t)]/2 < 1$.

\subsubsection{Pure States: $\alpha = 1, \beta = 0$ or $\alpha = 0, \beta = 1$}

For the initial state $\ket{\psi_0} = \ket{0, n+1}$ (i.e., $\alpha = 1$, $\beta = 0$):
\begin{equation}
F(t) = \frac{1 + D^4(t)}{2}.
\label{eq:fidelity_pure_0}
\end{equation}

Similarly, for $\ket{\psi_0} = \ket{1, n}$ (i.e., $\alpha = 0$, $\beta = 1$):
\begin{equation}
F(t) = \frac{1 + D^4(t)}{2}.
\label{eq:fidelity_pure_1}
\end{equation}

In both cases $R = \Re(\alpha\beta^*) = 0$, so the fidelity reduces to $(1+D^4)/2$, depending only on the decoherence factor and independent of the deterministic evolution phase $\Omega_0(t)$.

\subsection{Asymptotic Behavior}

\subsubsection{Short-time limit}

For small times $t \ll 1/\gamma$ (OU noise) or small $\Gamma t$ (white noise), the decoherence factor is approximately $D(t) \approx 1$, and thus:
\begin{equation}
F(t) \approx 1 - \mathcal{O}(t^2),
\label{eq:fidelity_short_time}
\end{equation}
indicating high fidelity for short evolution times.

\subsubsection{Long-time limit}

For long times $t \gg 1/\gamma$ (OU noise), the $e^{-\gamma t}$ term vanishes and the
variance grows linearly: $\mathcal{M}[\xi^2(t)] \approx \Gamma t/\gamma \cdot \gamma = \Gamma t$,
so the decoherence factor approaches
\begin{equation}
D_{\mathrm{OU}}(t) \to \exp\left[-\frac{(n+1)\Gamma t}{2}\right],
\label{eq:D_long_OU}
\end{equation}
which decays exponentially at the same rate as white noise, consistent with the
Markovian limit.

For white noise, $D_{\mathrm{white}}(t) = e^{-(n+1)\Gamma t/2}$ decays exponentially from the start. In the long-time limit:
\begin{equation}
F(t \to \infty) \to \frac{1}{2},
\label{eq:fidelity_long_time}
\end{equation}
representing complete loss of coherence and a maximally mixed state in the two-dimensional manifold.

\subsection{Fidelity and Quantum Gate Operations}

For quantum computing applications, specific evolution times are chosen to implement quantum gates. The most common choice is the quantum swap operation at time $t_{\pi/2}$ such that $\Omega_0(t_{\pi/2}) = \pi/2$, corresponding to:
\begin{equation}
t_{\pi/2} = \frac{\pi}{2g_0\sqrt{n+1}}.
\label{eq:t_pi_half}
\end{equation}

At this time, in the absence of noise, the ideal evolution would be:
\begin{equation}
\ket{\psi_{\mathrm{ideal}}(t_{\pi/2})} = -i\beta\ket{0, n+1} - i\alpha\ket{1, n}.
\label{eq:swap_ideal}
\end{equation}

The gate fidelity at this operation time is:
\begin{equation}
F_{\mathrm{gate}} = \frac{1 + D^4(t_{\pi/2})}{2} + 2R^2\bigl(1 - D^4(t_{\pi/2})\bigr).
\label{eq:gate_fidelity}
\end{equation}

For high-fidelity quantum operations, we require $F_{\mathrm{gate}} \approx 1$, which demands:
\begin{equation}
D^4(t_{\pi/2}) \approx 1 \quad \Rightarrow \quad \frac{(n+1)\Gamma t_{\pi/2}}{2} \ll 1.
\label{eq:high_fidelity_condition}
\end{equation}

This translates to the condition:
\begin{equation}
\frac{\Gamma}{g_0} \ll \frac{1}{\pi\sqrt{n+1}},
\label{eq:noise_coupling_ratio}
\end{equation}
indicating that the noise strength must be much smaller than the coupling strength, with the bound tightening for higher photon numbers.

\section{Numerical Results and Discussion}

In this section, we present comprehensive numerical results that illustrate the quantum dynamics of the qubit-cavity system under stochastic coupling fluctuations. We analyze the fidelity evolution, compare Markovian and non-Markovian regimes, examine photon number dependence, and provide phase diagrams characterizing the transition between different dynamical regimes.

\subsection{Fidelity Evolution under Varying Noise Intensity}

\begin{figure*}[htbp]
\centering
\includegraphics[width=0.95\textwidth]{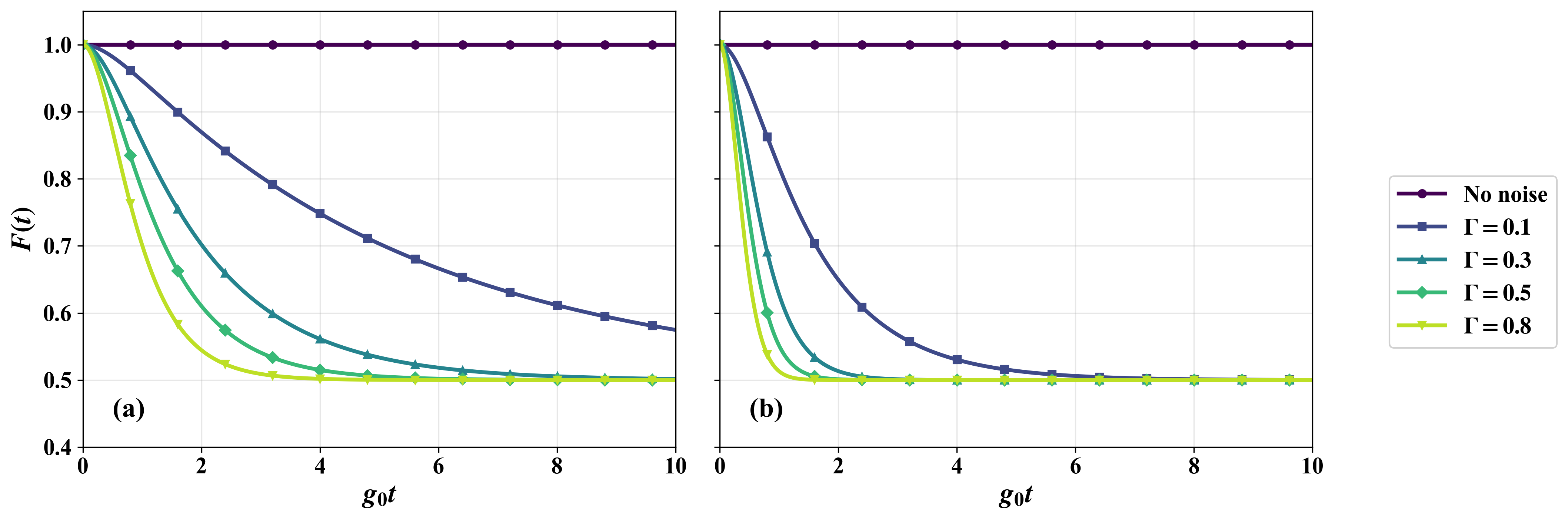}
\caption{Temporal evolution of fidelity $F(t)$ for initial cavity photon numbers
$n=0$ [panel~(a)] and $n=3$ [panel~(b)], with OU noise correlation rate
$\gamma = 2.0g_0$.  Curves correspond to noise intensities: no noise (purple),
$\Gamma = 0.1$ (blue), $\Gamma = 0.3$ (teal), $\Gamma = 0.5$ (green), and
$\Gamma = 0.8$ (yellow-green).  Higher noise intensities lead to faster fidelity
decay, with both configurations exhibiting asymptotic plateaus at long times.
The $n=3$ case shows enhanced sensitivity to noise during the initial transient
regime.  The legend, shared between both panels, is shown on the right.}
\label{fig:fidelity_noise_intensity}
\end{figure*}

Figure~\ref{fig:fidelity_noise_intensity} presents the temporal evolution of quantum state fidelity $F(t)$ under varying noise intensities for two distinct initial photon number configurations: $n=0$ [panel (a)] and $n=3$ [panel (b)], with correlation rate $\gamma = 2.0g_0$. In the absence of noise ($\Gamma = 0$, purple curves), both systems maintain near-perfect fidelity ($F(t) \approx 1.0$) throughout the evolution, demonstrating coherent dynamics with minimal degradation.

The introduction of noise parametrized by $\Gamma$ induces systematic fidelity degradation with several notable characteristics. All noise scenarios exhibit monotonically decreasing fidelity with time, consistent with irreversible decoherence processes. Higher noise intensities ($\Gamma = 0.8$, yellow-green) produce more rapid fidelity loss compared to moderate ($\Gamma = 0.3$, teal) and weak ($\Gamma = 0.1$, blue) noise regimes, and this trend is consistent across both photon number configurations. At extended times ($g_0 t \sim 10$), the fidelity curves exhibit plateauing behavior, approaching steady-state values that depend on both $\Gamma$ and $n$.

Comparing the two panels reveals the influence of initial cavity photon number on noise resilience. The $n=3$ configuration exhibits accelerated initial decay rates compared to $n=0$ for equivalent noise intensities, particularly evident in the $\Gamma = 0.5$ and $\Gamma = 0.8$ curves during the interval $0 < g_0 t < 2$. However, both systems converge to comparable asymptotic fidelities at long times, indicating that the initial photon number primarily affects transient dynamics rather than steady-state behavior. The parameter regime $\Gamma \leq 0.3$ maintains $F(t) > 0.6$ even at $g_0 t = 10$, suggesting potential operational thresholds for noise-tolerant quantum protocols.

\subsection{Markovian versus Non-Markovian Dynamics}

\begin{figure*}[htbp]
\centering
\includegraphics[width=0.7\textwidth]{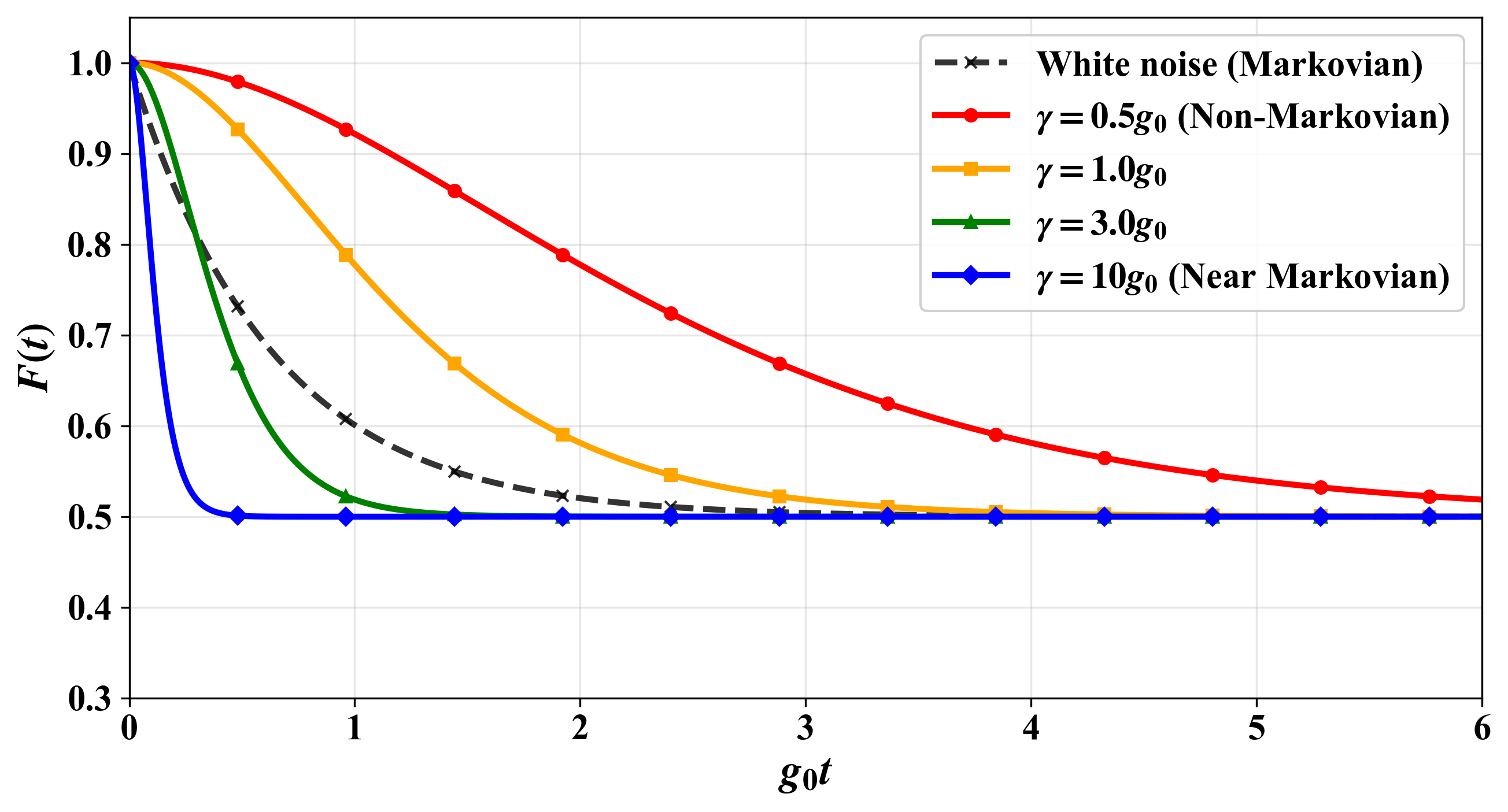}
\caption{Time evolution of fidelity $F(t)$ for OU noise with varying correlation
rates $\gamma$ compared to white noise, for $n=1$ and base noise intensity
$\Gamma_0 = 0.4$.  To enable a fair comparison across regimes, the OU noise
intensity is rescaled as $\Gamma = \Gamma_0\gamma$, so that the effective noise
strength $\Gamma/\gamma$ is held constant across all curves.  As $\gamma$
increases from $0.5g_0$ (red, strongly non-Markovian) to $10g_0$ (blue, near
Markovian), the dynamics transition toward white noise behavior (black dashed).
The x-axis range is $g_0 t \in [0,\,6]$.}
\label{fig:figure2_markovian_vs_nonmarkovian}
\end{figure*}

Figure~\ref{fig:figure2_markovian_vs_nonmarkovian} illustrates the transition from non-Markovian to Markovian dynamics in cavity quantum electrodynamics as the noise correlation rate $\gamma$ varies. The distinction between Markovian and non-Markovian dynamics is fundamentally determined by the noise correlation time $\tau_c = 1/\gamma$. In the Markovian limit, the correlation time becomes vanishingly short ($\gamma \to \infty$), meaning the noise at different times is essentially uncorrelated, approaching delta-correlated white noise. Conversely, in the non-Markovian regime, finite correlation times introduce memory effects where the noise at one instant influences future evolution.

Figure~\ref{fig:figure2_markovian_vs_nonmarkovian} presents fidelity evolution for four different correlation rates ranging from $\gamma = 0.5g_0$ (strongly non-Markovian, red) to $\gamma = 10g_0$ (near-Markovian, blue), alongside the white noise benchmark (black dashed). The $\gamma = 10g_0$ case nearly overlaps with white noise, confirming that at sufficiently large correlation rates the OU process converges to Markovian behavior. As $\gamma$ decreases, the curves progressively deviate from Markovian behavior, exhibiting slower initial decay and higher fidelities throughout the evolution window. The systematic variation of $\gamma$ from $0.5g_0$ to $10g_0$ maps the continuous transition between these regimes, revealing that the crossover occurs around $\gamma \sim 3g_0$.

\subsection{Photon Number Dependence of Decoherence}

\begin{figure*}[htbp]
\centering
\includegraphics[width=0.95\textwidth]{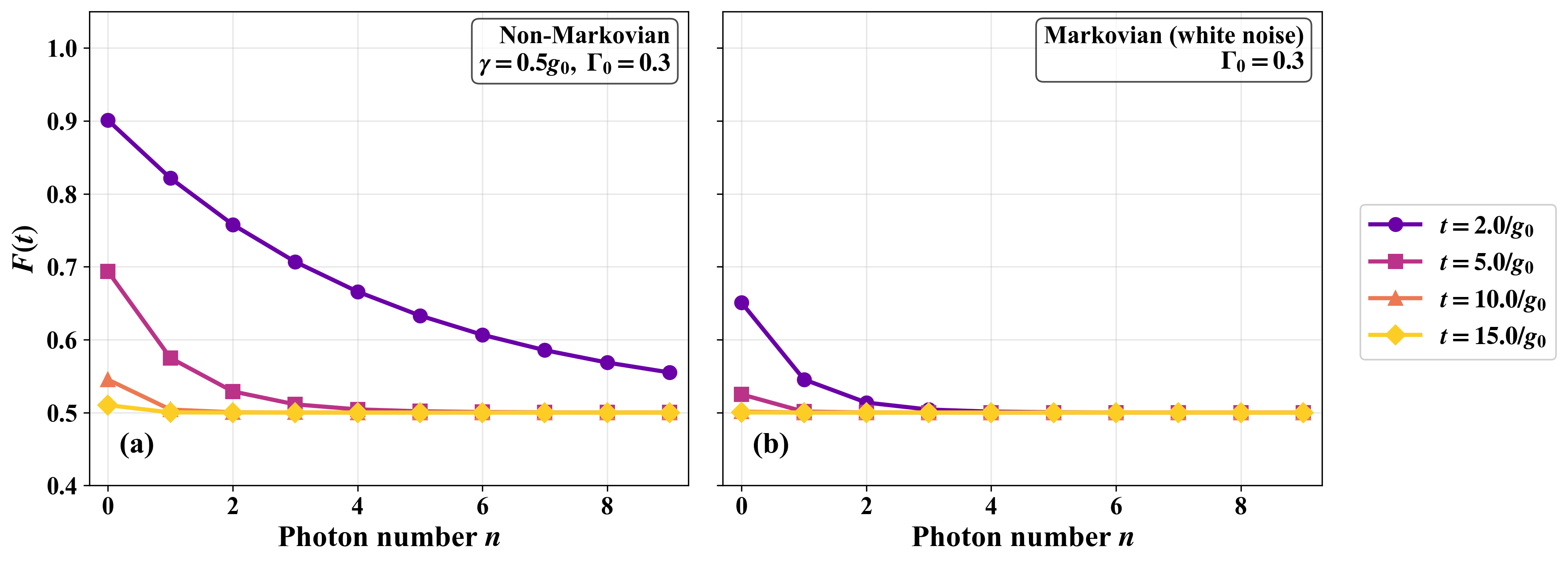}
\caption{Fidelity as a function of cavity photon number $n$ at fixed observation
times, comparing non-Markovian [panel~(a)] and Markovian [panel~(b)] dynamics.
Panel~(a): OU noise with $\gamma = 0.5g_0$ and rescaled intensity
$\Gamma = \Gamma_0\gamma = 0.15$ (base $\Gamma_0 = 0.3$).
Panel~(b): Markovian white noise with $\Gamma_0 = 0.3$.
Curves shown for $t = 2.0/g_0$ (dark purple), $t = 5.0/g_0$ (magenta),
$t = 10.0/g_0$ (orange-red), and $t = 15.0/g_0$ (yellow).
Fidelity decreases monotonically with photon number due to the $(n+1)$ scaling
in the decoherence factor, with the steepest drop from $n=0$ to $n=1$.}
\label{fig:photon_dependence}
\end{figure*}

Figure~\ref{fig:photon_dependence} illustrates how the fidelity depends on the cavity photon number $n$ for fixed observation times. In both panels, the fidelity decreases monotonically with increasing photon number for all observation times. This universal trend reflects the $(n+1)$ scaling in the decoherence factor, where higher photon numbers experience amplified dephasing through the enhanced coupling strength $g_0\sqrt{n+1}$ in the Jaynes-Cummings interaction. The decline is particularly steep for the transition from $n = 0$ to $n = 1$, where the effective coupling increases by approximately 41\%.

A key observation from both panels is that the photon number dependence becomes progressively weaker at later times. At $t = 2.0/g_0$, there is substantial variation in fidelity across the photon number range, with a drop of approximately 0.15--0.20. However, by $t = 15.0/g_0$, all curves have nearly converged to the asymptotic plateau, showing minimal photon number dependence.

\subsection{Phase Diagram of Dynamical Regimes}

\begin{figure*}[htbp]
\centering
\includegraphics[width=0.95\textwidth]{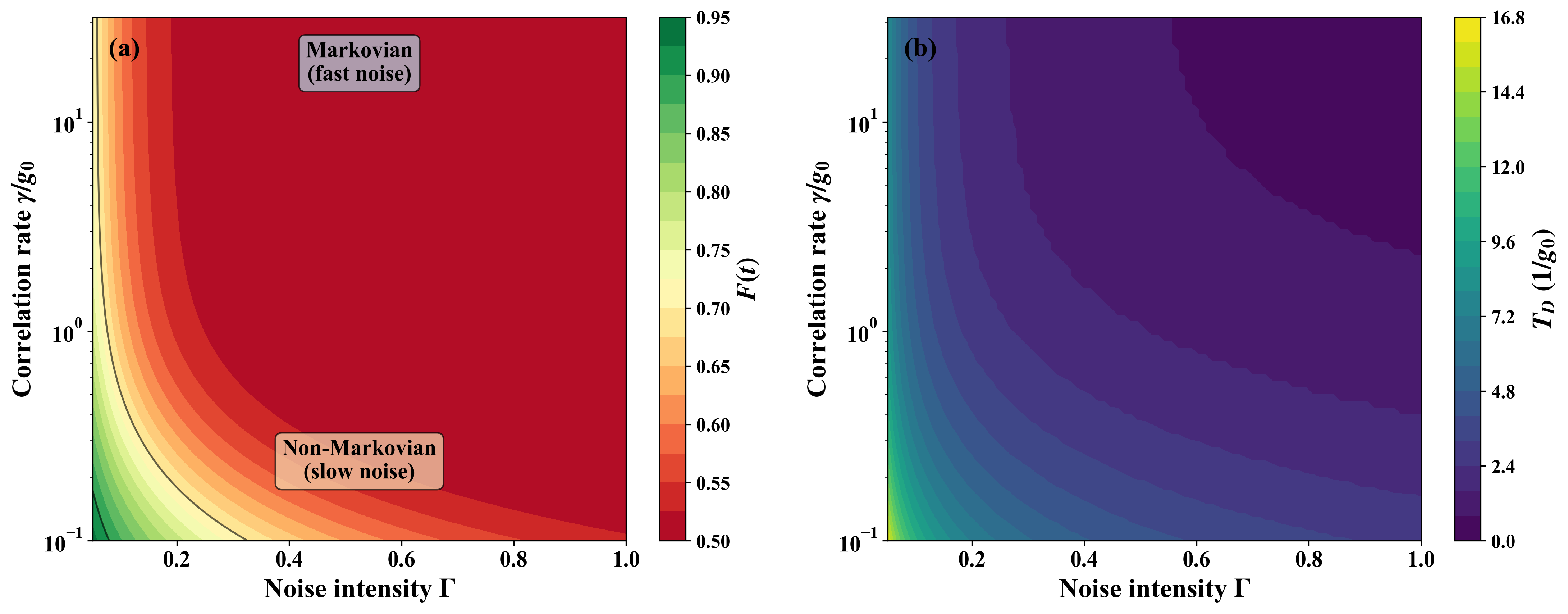}
\caption{\textbf{Phase diagram of Markovian to non-Markovian transition} for $n=1$.
Panel~(a): fidelity map $F(\Gamma,\gamma)$ at fixed time $t = 4.0/g_0$.
Contour lines at $F = 0.5,\,0.7,\,0.9$ delineate regions of low,
moderate, and high fidelity.  The non-Markovian regime (bottom, $\gamma/g_0 \lesssim
1$) maintains substantially higher fidelity than the Markovian regime (top,
$\gamma/g_0 \gtrsim 10$) at equal $\Gamma$.
Panel~(b): decoherence time $T_D$, defined as the time for fidelity to drop to
$F = 0.6$, across the same parameter space.  In the non-Markovian regime $T_D$ can
reach nearly five times the Markovian value at intermediate noise intensity
($\Gamma \sim 0.4$): $T_D \approx 4.9/g_0$ vs.\ $1.0/g_0$.}
\label{fig:transition_map}
\end{figure*}

Figure~\ref{fig:transition_map} presents two-dimensional phase diagrams that map out the fidelity landscape and decoherence timescales as functions of both the noise intensity $\Gamma$ and the correlation rate $\gamma/g_0$. The left panel shows a fidelity map at fixed time $t = 4.0/g_0$ for photon number $n=1$, while the right panel displays the decoherence time $T_D$ defined as the time required for fidelity to drop to $F = 0.6$.

The fidelity map in panel~(a) reveals several structural features. In the bottom region where $\gamma/g_0 \lesssim 1$ (non-Markovian regime), the fidelity is substantially higher than in the corresponding Markovian region at the same $\Gamma$. The non-Markovian regime maintains $F > 0.7$ up to $\Gamma \approx 0.3$, while the Markovian regime falls below $F = 0.7$ already for $\Gamma \gtrsim 0.05$. The decoherence time map in panel~(b) corroborates these trends quantitatively. The most striking contrast occurs at intermediate noise intensities: at $\Gamma = 0.4$, the non-Markovian regime gives $T_D \approx 4.9/g_0$, while the Markovian limit gives $T_D \approx 1.0/g_0$---nearly a five-fold difference at equal noise intensity.

\subsection{Non-Markovian Dynamics: Enhanced Coherence Preservation}

\begin{figure*}[htbp]
\centering
\includegraphics[width=0.7\textwidth]{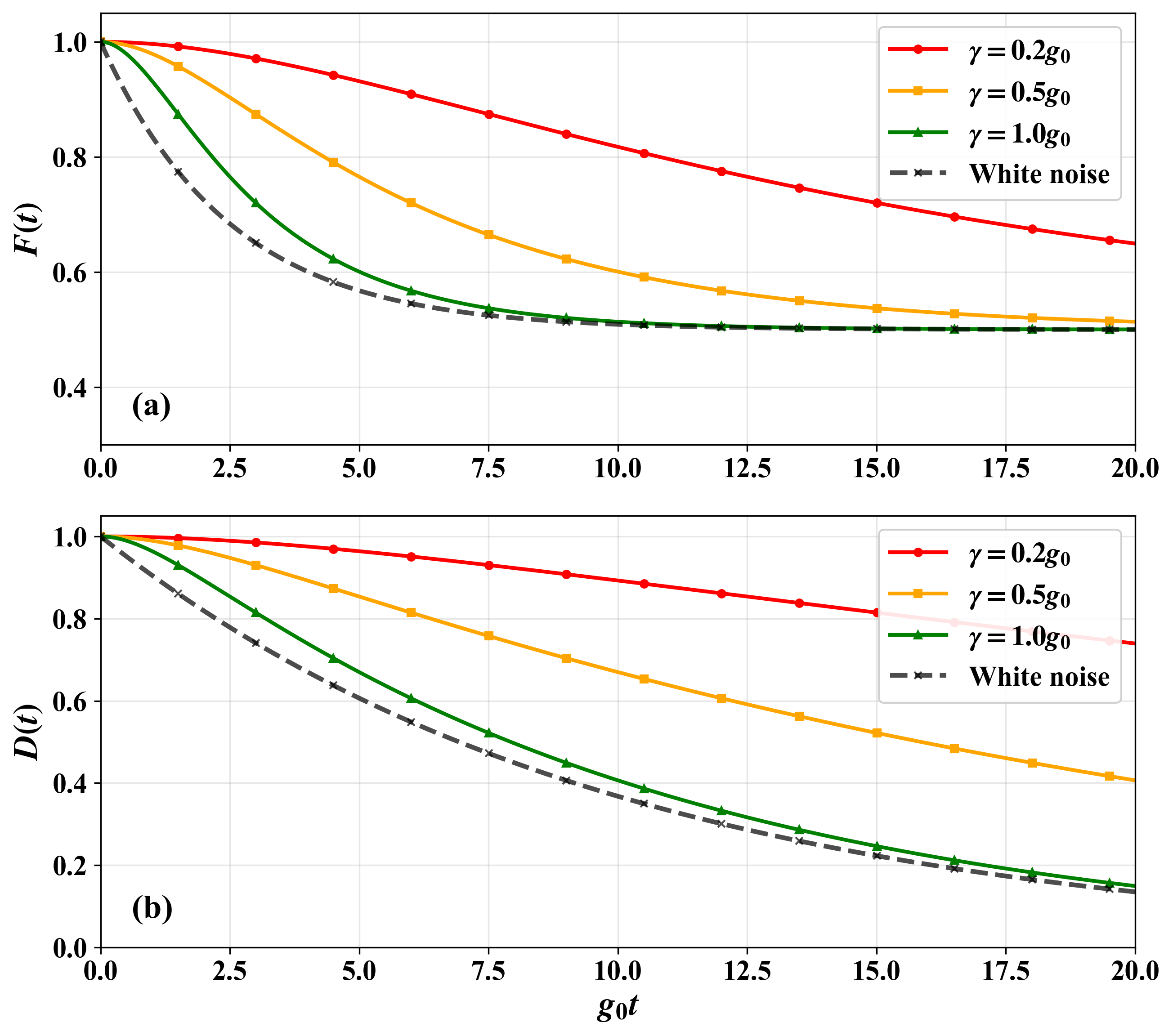}
\caption{\textbf{Enhanced coherence preservation in the non-Markovian regime},
for $n=0$ and base noise intensity $\Gamma_0=0.2$.  The OU noise intensity is
rescaled as $\Gamma = \Gamma_0\gamma$ for each curve.
Panel~(a): fidelity $F(t)$ for $\gamma = 0.2g_0$ (red), $0.5g_0$ (orange),
$1.0g_0$ (green), and white noise (black dashed).  Slower noise (smaller $\gamma$)
maintains higher fidelity throughout the evolution.
Panel~(b): corresponding decoherence factor $D(t)$, showing non-exponential
decay in the non-Markovian regime and the hierarchy
$D_{\gamma=0.2} > D_{\gamma=0.5} > D_{\gamma=1.0} > D_{\text{white}}$.}
\label{fig:oscillatory}
\end{figure*}

Figure~\ref{fig:oscillatory} focuses on the non-Markovian regime, especially where the decay rate is much smaller than the coupling strength ($\gamma \ll g_0$). The figure reveals enhanced coherence preservation and characteristically slower decoherence compared to Markovian (white noise) dynamics.

The fidelity curves reveal clear differences between non-Markovian and Markovian dynamics. For the slowest noise ($\gamma = 0.2g_0$, red curve), the fidelity decays significantly more slowly than white noise throughout the entire evolution. As the correlation decay rate $\gamma$ increases, the protection diminishes. At early times ($g_0 t = 5$), the slowest noise maintains fidelity above 0.8 while white noise drops to approximately 0.6, demonstrating substantial protection from long noise correlations. Panel~(b) quantifies the coherence preservation more directly through $D(t)$: at $g_0 t = 20$, the slowest noise achieves $D_{\gamma=0.2} \approx 0.22$ compared to $D_{\text{white}} \approx 0.13$---a nearly two-fold enhancement.

\section{Conclusions}

We have analytically studied the dynamics of a single qubit coupled to a cavity field under stochastic coupling fluctuations, modeling the noise using both white noise and Ornstein-Uhlenbeck processes. Our analysis reveals that the state fidelity is given by the general 
expression $F(t) = \frac{1+D^4(t)}{2} + 2R^2(1-D^4(t))$, where 
$R = \Re(\alpha\beta^*)$ characterizes the initial state. For states 
with $R = 0$ (including all pure states $\alpha=1$ or $\beta=1$), this 
simplifies to $F(t) = [1+D^4(t)]/2$, independent of the deterministic 
evolution phase. This remarkably simple result provides a direct connection between microscopic noise properties and macroscopic fidelity loss.

The transition between dynamical regimes is controlled by the dimensionless ratio $\gamma/g_0$. In the Markovian limit ($\gamma/g_0 \gtrsim 10$), the dynamics converge to white noise behavior with smooth exponential decay. In the non-Markovian regime ($\gamma/g_0 \ll 1$), long noise correlations produce non-exponential decay completely absent in Markovian systems. The crossover occurs around $\gamma/g_0 \sim 1$-$3$, providing a practical criterion for selecting appropriate theoretical frameworks.

Decoherence scales as $(n+1)$ with photon number, making higher Fock states progressively more vulnerable to noise. The fidelity decreases monotonically with increasing $n$, with the steepest drop from $n = 0$ to $n = 1$ where the effective coupling increases by approximately 41\%.

Counterintuitively, correlated noise with long correlation times is less detrimental to quantum coherence than uncorrelated noise of equal intensity. Our phase diagrams reveal that high fidelity ($F > 0.7$ at $t = 10/g_0$) requires low noise intensity ($\Gamma \lesssim 0.4$), with the non-Markovian regime providing additional protection through memory effects. The decoherence time $T_D$ exceeds $20/g_0$ for weak noise but drops to $T_D \lesssim 5/g_0$ under strong noise ($\Gamma \gtrsim 0.8$), regardless of regime.

These results provide both fundamental insights and practical guidance. Understanding the noise correlation time $1/\gamma$ relative to the system timescale $g_0^{-1}$ is as important as measuring noise intensity for predicting quantum coherence behavior.



\section*{Data Availability}
No new data were created or analyzed in this study.

\appendix

\section{Detailed derivation of Time Evolution}
\label{app:derivation}

The time evolution operator in the interaction picture is given by
\begin{equation}
U(t) = \mathcal{T} \exp\left[ -i \int_0^t H_{\mathrm{int}}(\tau) \, d\tau \right],
\label{eq:evolution_operator_app}
\end{equation}
where $\mathcal{T}$ is the time-ordering operator. Since $H_{\mathrm{int}}(t)$ and $H_{\mathrm{int}}(s)$ commute for all $t$ and $s$ (as $g(t)$ is a scalar function), we can drop the time-ordering:
\begin{equation}
U(t) = \exp\left[ -i \int_0^t g(\tau) \, d\tau \left( a^{\dagger} \sigma_{-} + a \sigma_{+} \right) \right].
\label{eq:evolution_simplified}
\end{equation}

Define the effective coupling parameter:
\begin{equation}
G(t) = \int_0^t g(\tau) \, d\tau = g_0 t + \int_0^t \eta(\tau) \, d\tau.
\label{eq:effective_coupling_app}
\end{equation}

Then the evolution operator becomes
\begin{equation}
U(t) = \exp\left[ -i G(t) \left( a^{\dagger} \sigma_{-} + a \sigma_{+} \right) \right].
\label{eq:U_G_app}
\end{equation}

The JC model interaction couples pairs of states with the same total excitation number. For a system with $n$ photons in the cavity, the interaction couples the manifold $\{\ket{0, n+1}, \ket{1, n}\}$. We consider an initial state of the form
\begin{equation}
\ket{\psi_0} = \alpha \ket{0, n+1} + \beta \ket{1, n},
\label{eq:initial_state_app}
\end{equation}
where $\alpha$ and $\beta$ are complex coefficients with $|\alpha|^2 + |\beta|^2 = 1$.

The interaction Hamiltonian in this subspace can be written in matrix form:
\begin{equation}
H_{\mathrm{int}}(t) = g(t) \sqrt{n+1} 
\begin{pmatrix}
0 & 1 \\
1 & 0
\end{pmatrix} = g(t) \sqrt{n+1} \, \sigma_x,
\label{eq:H_matrix_app}
\end{equation}
where $\sigma_x$ is the Pauli-$x$ matrix in the $\{\ket{0, n+1}, \ket{1, n}\}$ basis.

The evolution operator in this subspace is
\begin{equation}
\begin{split}
U(t) &= \exp\left[-i G(t) \sqrt{n+1} \, \sigma_x \right] \\
&= \cos\left(G(t)\sqrt{n+1}\right) \mathbf{1} - i \sin\left(G(t)\sqrt{n+1}\right) \sigma_x,
\end{split}
\label{eq:U_matrix}
\end{equation}
where $\mathbf{1}$ is the $2 \times 2$ identity matrix. Explicitly:
\begin{equation}
U(t) = 
\begin{pmatrix}
\cos(\Omega) & -i \sin(\Omega) \\
-i \sin(\Omega) & \cos(\Omega)
\end{pmatrix},
\label{eq:U_explicit_app}
\end{equation}
where $\Omega = G(t)\sqrt{n+1}$.

For the initial state given in Eq.~\eqref{eq:initial_state_app}, the evolved state is:
\begin{equation}
\begin{split}
\ket{\psi(t)} = & U(t) \ket{\psi_0}\\
=&\left[\alpha \cos(\Omega) - i \beta \sin(\Omega)\right] \ket{0, n+1} \\
& + \left[\beta \cos(\Omega) - i \alpha \sin(\Omega)\right] \ket{1, n}.
\end{split}
\label{eq:evolved_state_app}
\end{equation}
Due to the stochastic nature of $g(t)$, we must perform ensemble averaging. The density matrix of the system is:
\begin{equation}
\rho(t) = \mathcal{M}[U(t) \ket{\psi_0}\bra{\psi_0} U^\dagger(t)].
\label{eq:density_matrix_app}
\end{equation}
The key quantity to compute is $\mathcal{M}[\cos(\Omega)]$ and $\mathcal{M}[\sin(\Omega)]$ where $\Omega = G(t)\sqrt{n+1}$.

For OU noise, $G(t) = g_0 t + \xi(t)$ where $\xi(t) = \int_0^t \eta(\tau) d\tau$ is a Gaussian random variable with:
\begin{align}
\mathcal{M}[\xi(t)] &= 0, \\
\mathcal{M}[\xi^2(t)] &= \int_0^t \int_0^t \mathcal{M}[\eta(\tau)\eta(s)] \, d\tau ds.
\end{align}

For OU noise with correlation function Eq.~\eqref{eq:ou_correlation}:
\begin{equation}
\mathcal{M}[\xi^2(t)] = \frac{\Gamma}{\gamma}\left[\gamma t - 1 + e^{-\gamma t}\right].
\label{eq:xi_variance_OU}
\end{equation}

For white noise (limit $\gamma \to \infty$):
\begin{equation}
\mathcal{M}[\xi^2(t)] = \Gamma t.
\label{eq:xi_variance_white}
\end{equation}

Since $\xi(t)$ is Gaussian, we can use:
\begin{align}
\mathcal{M}[e^{i k \xi}] &= e^{-k^2 \mathcal{M}[\xi^2]/2}, \\
\mathcal{M}[\cos(k\xi)] &= e^{-k^2 \mathcal{M}[\xi^2]/2}, \\
\mathcal{M}[\sin(k\xi)] &= 0.
\end{align}

Therefore:
\begin{align}
\mathcal{M}[\cos(\Omega)] &= \cos(g_0 t\sqrt{n+1}) \, e^{-(n+1)\mathcal{M}[\xi^2(t)]/2}, \label{eq:cos_avg}\\
\mathcal{M}[\sin(\Omega)] &= \sin(g_0 t\sqrt{n+1}) \, e^{-(n+1)\mathcal{M}[\xi^2(t)]/2}. \label{eq:sin_avg}
\end{align}

Define the decoherence factor:
\begin{equation}
D(t) = e^{-(n+1)\mathcal{M}[\xi^2(t)]/2}.
\label{eq:decoherence_factor_app}
\end{equation}

For OU noise:
\begin{equation}
D_{\mathrm{OU}}(t) = \exp\left[-\frac{(n+1)\Gamma}{2\gamma}\left(\gamma t - 1 + e^{-\gamma t}\right)\right].
\label{eq:D_OU_app}
\end{equation}

For white noise:
\begin{equation}
D_{\mathrm{white}}(t) = e^{-(n+1)\Gamma t/2}.
\label{eq:D_white_app}
\end{equation}

\subsection{Noise-averaged density matrix}

We now evaluate the noise-averaged density matrix
$\rho(t) = \mathcal{M}\!\big[\,|\psi(t)\rangle\langle\psi(t)|\,\big]$
for the evolved state in Eq.~\eqref{eq:evolved_state_app}. Working in the two-dimensional JC manifold spanned by $\{|0,n+1\rangle, |1,n\rangle\}$, the density matrix elements before averaging over the noise are:
\begin{small}
\begin{align}
\rho_{00}(t) &= |\alpha|^2 \cos^2\Omega + |\beta|^2 \sin^2\Omega \nonumber \\
      &- \Im\!\left(\alpha\beta^*\right)\sin 2\Omega, \label{eq:rho00_unavg}\\[2mm]
\rho_{11}(t) &= |\alpha|^2 \sin^2\Omega + |\beta|^2 \cos^2\Omega \nonumber \\
      &+ \Im\!\left(\alpha\beta^*\right)\sin 2\Omega, \label{eq:rho11_unavg}\\[2mm]
\rho_{01}(t) &= \alpha\beta^* \cos^2\Omega + \alpha^*\beta \sin^2\Omega \nonumber \\
   &+ i\big(|\alpha|^2 - |\beta|^2\big)\cos\Omega\sin\Omega, \label{eq:rho01_unavg}\\[2mm]
\rho_{10}(t) &= \rho_{01}^*(t).
\end{align}
\end{small}

\subsection{Gaussian averaging of trigonometric terms}

It is convenient to separate the deterministic and stochastic parts of $\Omega(t)$. Using
\begin{align}
\Omega(t) &= \sqrt{n+1}\,G(t) \nonumber \\
&= \Omega_0(t) + \sqrt{n+1}\,\xi(t),
\end{align}
with $\Omega_0(t) = g_0 t\sqrt{n+1}$ and $\xi(t) = \int_0^t \eta(\tau)\,d\tau$.

For the quadratic combinations entering Eqs.~\eqref{eq:rho00_unavg}--\eqref{eq:rho01_unavg} we use the identities for double-angle combinations. Since $2\Omega(t) = 2\Omega_0(t) + 2\sqrt{n+1}\,\xi(t)$ and $\xi(t)$ is Gaussian with zero mean:
\begin{align}
\mathcal{M}[\cos 2\Omega] 
&= \cos(2\Omega_0(t))\,D^4(t), \label{eq:cos2Omega_avg}\\[1mm]
\mathcal{M}[\sin 2\Omega] 
&= \sin(2\Omega_0(t))\,D^4(t). \label{eq:sin2Omega_avg}
\end{align}
From these, the required averages follow:
\begin{align}
\mathcal{M}[\cos^2\Omega] 
&= \frac{1}{2}\Big(1 + \cos(2\Omega_0(t))\,D^4(t)\Big), \label{eq:cos2_avg}\\
\mathcal{M}[\sin^2\Omega] 
&= \frac{1}{2}\Big(1 - \cos(2\Omega_0(t))\,D^4(t)\Big), \label{eq:sin2_avg}\\
\mathcal{M}[\cos\Omega\sin\Omega] 
&= \frac{1}{2}\sin(2\Omega_0(t))\,D^4(t). \label{eq:cossin_avg}
\end{align}

\subsection{Explicit form of the noise-averaged density matrix}

Define
\begin{align}
\Delta &= |\alpha|^2 - |\beta|^2, \nonumber \\
R &= \Re(\alpha\beta^*), \nonumber \\
I &= \Im(\alpha\beta^*),
\end{align}
and introduce the shorthand
\begin{align}
C(t) &= \cos\big(2\Omega_0(t)\big)\,D^4(t), \\
S(t) &= \sin\big(2\Omega_0(t)\big)\,D^4(t).
\end{align}
Assuming the initial qubit state is normalized, $|\alpha|^2 + |\beta|^2 = 1$, inserting Eqs.~\eqref{eq:cos2_avg}--\eqref{eq:cossin_avg} into Eqs.~\eqref{eq:rho00_unavg}--\eqref{eq:rho01_unavg} yields
\begin{footnotesize}
\begin{align}
\rho_{00}(t)
&= \frac{1}{2}
  + \frac{\Delta}{2} C(t)
  - I\, S(t), \label{eq:rho00_final_app}\\[2mm]
\rho_{11}(t)
&= \frac{1}{2}
  - \frac{\Delta}{2} C(t)
  + I\, S(t), \label{eq:rho11_final_app}\\[2mm]
\rho_{01}(t)
&= R + i I\, C(t)
   + i\frac{\Delta}{2} S(t), \label{eq:rho01_final_app}\\[2mm]
\rho_{10}(t) &= \rho_{01}^*(t). \label{eq:rho10_final_app}
\end{align}
\end{footnotesize}

Thus, the noise-averaged density matrix in the $\{|0,n+1\rangle,|1,n\rangle\}$ manifold takes the compact form
\begin{small}
\begin{equation}
\rho(t) =
\begin{pmatrix}
\frac{1}{2} + \frac{\Delta}{2} C(t) - I S(t)
&
R + i I C(t) + i\frac{\Delta}{2} S(t) \\[1mm]
R - i I C(t) - i\frac{\Delta}{2} S(t)
&
\frac{1}{2} - \frac{\Delta}{2} C(t) + I S(t)
\end{pmatrix}.
\label{eq:rho_matrix_final_app}
\end{equation}
\end{small}

\subsection{Reduced qubit state}

If we are interested only in the qubit, we trace out the cavity mode. In the present single-manifold description, the reduced qubit density matrix is diagonal in the energy basis because $\langle n+1|n\rangle = 0$:
\begin{equation}
\rho_{\mathrm{q}}(t) = \mathrm{Tr}_{\mathrm{cav}}[\rho(t)]
= \rho_{00}(t)\,|0\rangle\langle 0|
 + \rho_{11}(t)\,|1\rangle\langle 1|.
\label{eq:rho_qubit_app}
\end{equation}

\section{Fidelity Analysis}

\subsection{Fidelity Definition}

For a mixed state described by density matrix $\rho$ and a pure target state $\ket{\psi_{\mathrm{target}}}$, the fidelity is defined as
\begin{equation}
F = \bra{\psi_{\mathrm{target}}} \rho \ket{\psi_{\mathrm{target}}}.
\label{eq:fidelity_def_app}
\end{equation}

\subsection{Ideal Target State}

For purely deterministic evolution with coupling strength $g_0$, the evolution operator is
\begin{equation}
U_0(t) = \exp\left[-i g_0 t \sqrt{n+1} \, \sigma_x \right],
\end{equation}
and the ideal evolved state is
\begin{equation}
\ket{\psi_{\mathrm{ideal}}(t)} = U_0(t) \ket{\psi_0}.
\label{eq:ideal_state_app}
\end{equation}

Let $\Omega_0(t) = g_0 t \sqrt{n+1}$. Then the ideal state is
\begin{equation}
\begin{split}
\ket{\psi_{\mathrm{ideal}}(t)} 
&= \left[\alpha \cos(\Omega_0) - i \beta \sin(\Omega_0)\right] \ket{0, n+1} \\
& \quad + \left[\beta \cos(\Omega_0) - i \alpha \sin(\Omega_0)\right] \ket{1, n}.
\end{split}
\label{eq:ideal_evolved_app}
\end{equation}

\subsection{Fidelity Calculation}

The fidelity between the noise-averaged density matrix $\rho(t)$ and the ideal state is
\begin{equation}
F(t) = \bra{\psi_{\mathrm{ideal}}(t)} \rho(t) \ket{\psi_{\mathrm{ideal}}(t)}.
\label{eq:fidelity_app}
\end{equation}

\subsection{Evaluation of Fidelity Components}

We first compute $|c_0^{\mathrm{ideal}}|^2$ and $|c_1^{\mathrm{ideal}}|^2$:
\begin{align}
|c_0^{\mathrm{ideal}}|^2 &= |\alpha|^2 \cos^2(\Omega_0) \nonumber \\
&+ |\beta|^2 \sin^2(\Omega_0) - \Im(\alpha\beta^*) \sin(2\Omega_0), \label{eq:c0_ideal_sq}\\
|c_1^{\mathrm{ideal}}|^2 &= |\alpha|^2 \sin^2(\Omega_0) + |\beta|^2 \cos^2(\Omega_0) \nonumber \\
&+ \Im(\alpha\beta^*) \sin(2\Omega_0). \label{eq:c1_ideal_sq}
\end{align}

Next, we compute the off-diagonal contribution:
\begin{align}
(c_0^{\mathrm{ideal}})^* c_1^{\mathrm{ideal}} 
&= (\alpha^* \cos(\Omega_0) + i \beta^* \sin(\Omega_0))\cdot \nonumber \\
&\quad (\beta \cos(\Omega_0) - i \alpha \sin(\Omega_0)) \nonumber \\
&= \alpha^*\beta \cos^2(\Omega_0) + \alpha\beta^* \sin^2(\Omega_0) \nonumber \\
& \quad + i(|\alpha|^2 - |\beta|^2) \cos(\Omega_0)\sin(\Omega_0).
\label{eq:c0c1_ideal}
\end{align}

Using $\alpha^*\beta = R - iI$ and $\alpha\beta^* = R + iI$, along with the notation $\Delta = |\alpha|^2 - |\beta|^2$:
\begin{equation}
(c_0^{\mathrm{ideal}})^* c_1^{\mathrm{ideal}} = R - i I \cos(2\Omega_0) - i \frac{\Delta}{2} \sin(2\Omega_0).
\label{eq:c0c1_simplified}
\end{equation}

Similarly:
\begin{equation}
(c_1^{\mathrm{ideal}})^* c_0^{\mathrm{ideal}} = R + i I \cos(2\Omega_0) + i \frac{\Delta}{2} \sin(2\Omega_0).
\label{eq:c1c0_simplified}
\end{equation}

\subsection{Explicit Fidelity Expression}

After algebraic manipulation using Eqs.~\eqref{eq:c0_ideal_sq}--\eqref{eq:c1c0_simplified} and the density matrix elements from Eqs.~\eqref{eq:rho00_final_app}--\eqref{eq:rho10_final_app}, the $I\Delta$ cross-terms cancel identically between the diagonal and off-diagonal contributions. Collecting the remaining terms and applying $\cos(2\Omega_0)C(t)+\sin(2\Omega_0)S(t)=D^4(t)$ together with $\Delta^2/2 + 2I^2 = 1/2 - 2R^2$ (which follows from $|\alpha|^2+|\beta|^2=1$ and $|\alpha\beta^*|^2 = R^2+I^2$), the fidelity simplifies to:
\begin{equation}
F(t) = \frac{1 + D^4(t)}{2} + 2R^2 \big(1 - D^4(t)\big),
\label{eq:fidelity_compact_app}
\end{equation}
where $R = \Re(\alpha\beta^*)$. This can be rewritten in alternative forms:
\begin{equation}
F(t) = \frac{1 + D^4(t)}{2}\bigl(1 - 4R^2\bigr) + 2R^2,
\label{eq:fidelity_alternative}
\end{equation}
or equivalently,
\begin{equation}
F(t) = 1 - \bigl(1 - D^4(t)\bigr)\bigl(1 - 2R^2\bigr).
\label{eq:fidelity_alternative2}
\end{equation}

\subsection{Special Cases}

\subsubsection{Equal Superposition: $|\alpha| = |\beta| = 1/\sqrt{2}$ (real coefficients)}

For an initially balanced superposition with real coefficients $\alpha = \beta = 1/\sqrt{2}$, we have $R = 1/2$, so $2R^2 = 1/2$ and the fidelity becomes:
\begin{equation}
F(t) = \frac{1 + D^4(t)}{2} + \frac{1 - D^4(t)}{2} = 1.
\label{eq:fidelity_equal_app}
\end{equation}

\subsubsection{Pure States: $\alpha = 1$ or $\beta = 1$}

For the initial state $\ket{\psi_0} = \ket{0, n+1}$ (i.e., $\alpha = 1$, $\beta = 0$): here $R=0$, so
\begin{equation}
F(t) = \frac{1 + D^4(t)}{2}.
\label{eq:fidelity_pure_0_app}
\end{equation}

\subsection{Asymptotic Behavior}

For small times $t \ll 1/\gamma$ (OU noise) or small $\Gamma t$ (white noise):
\begin{equation}
F(t) \approx 1 - \mathcal{O}(t^2).
\label{eq:fidelity_short_time_app}
\end{equation}

For long times $t \gg 1/\gamma$ (OU noise):
\begin{equation}
D_{\mathrm{OU}}(t) \to \exp\left[-\frac{(n+1)\Gamma t}{2}\right],
\label{eq:D_long_OU_app}
\end{equation}
and in the long-time limit for both noise models:
\begin{equation}
F(t \to \infty) \to \frac{1}{2}.
\label{eq:fidelity_long_time_app}
\end{equation}

\subsection{Fidelity and Quantum Gate Operations}

The most common choice is the quantum swap operation at time $t_{\pi/2}$ such that $\Omega_0(t_{\pi/2}) = \pi/2$:
\begin{equation}
t_{\pi/2} = \frac{\pi}{2g_0\sqrt{n+1}}.
\label{eq:t_pi_half_app}
\end{equation}
The gate fidelity at this operation time is:
\begin{small}
\begin{equation}
F_{\mathrm{gate}} = F(t_{\pi/2}) 
= \frac{1 + D^4(t_{\pi/2})}{2} 
+ 2R^2\bigl(1 - D^4(t_{\pi/2})\bigr),
\label{eq:gate_fidelity_app}
\end{equation}
\end{small}
where $R = \Re(\alpha\beta^*)$ depends on the initial state.
For initial states with $R = 0$, this reduces to
\begin{equation}
F_{\mathrm{gate}} = \frac{1 + D^4(t_{\pi/2})}{2}.
\end{equation}

For high-fidelity quantum operations, the condition reads:
\begin{equation}
\frac{\Gamma}{g_0} \ll \frac{1}{\pi\sqrt{n+1}}.
\label{eq:noise_coupling_ratio_app}
\end{equation}

\renewcommand\refname{Reference}
\bibliographystyle{apsrev4-2}
\bibliography{bib_fang.bib}

\end{document}